\documentclass[aip,reprint]{revtex4-1}
\usepackage{bm}
\usepackage{graphicx}
\usepackage{bm}
\usepackage[dvips]{color}
\usepackage{natbib}
\begin{document}

\title{Study of spin-phonon coupling and magnetic field induced spin reorientation in polycrystalline multiferroic $GdFeO_3$}


\author{Anjali Panchwanee}
\affiliation{UGC-DAE Consortium for Scientific Research, University Campus, Khandwa Road, Indore 452001, India.}

\author{V. Raghavendra Reddy} 
\email{varimalla@yahoo.com; vrreddy@csr.res.in}
\affiliation{UGC-DAE Consortium for Scientific Research, University Campus, Khandwa Road, Indore 452001, India.}

\author{Ajay Gupta} 
\affiliation{Amity Center for Spintronic Materials, Amity University, Noida 201303, India.}

\author{V. G. Sathe}
\affiliation{UGC-DAE Consortium for Scientific Research, University Campus, Khandwa Road, Indore 452001, India.}

\begin{abstract}

The present work reports the preparation of polycrystalline multiferroic $GdFeO_3$ (GdFO) and characterization with x-ray diffraction (XRD), magnetization, temperature dependent Raman spectroscopy, temperature and magnetic field dependent $^{57}Fe$ M$\ddot{o}$ssbauer spectroscopy measurements. The sample is found to be phase pure from Rietveld refinement of XRD pattern. The M$\ddot{o}$ssbauer spectra measured in the presence of external magnetic field show the signatures of field induced spin reorientation transition, which are corroborated by magnetization measurements. From the temperature dependent variation of internal hyperfine field, N$\grave{e}$el transition temperature ($T_{N,Fe}$) of 672.5$\pm$0.2 K and critical exponent ($\beta$) of 0.333$\pm$0.003 is obtained. Temperature dependent (300 - 760 K) Raman spectroscopy measurements show the signatures of spin-phonon coupling  and local structural re-arrangement across $T_{N,Fe}$.       
 
\end{abstract}



\maketitle \section{Introduction}

In recent years, multiferroic / magneto-electric (ME) materials are the subject of active research because of rich physics they exhibit and also various potential applications \cite{Prellier, Cheong, RRamesh, CNRReview}. Multiferroic materials show electric and magnetic ordering simultaneously with a coupling between the two order parameters and therefore offer an extra degree of freedom in device design \cite{Prellier, Cheong, RRamesh, CNRReview}.  In order to have considerable ME coupling it is reported that magnetism induced polarization is more favorable \cite{Prellier, Cheong, RRamesh}. Rare-earth ortho-ferrites ($RFeO_3$ with R being rare-earth ion) comes under this category of multiferroic materials \cite{CNRReview}. 

In general, $RFeO_3$ compounds crystallize in orthorhombic distorted ($ABO_3$) perovskite structure with centro-symmetric space group $\textit{Pbnm}$ and exhibits $Fe^{3+}$ - $Fe^{3+}$, $R^{3+}$- $Fe^{3+}$ and $R^{3+}$- $R^{3+}$ magnetic exchange interactions resulting in various magnetic transitions viz., paramagnetic to antiferromagnetic (AFM) ordering of $Fe^{3+}$ ions (N$\grave{e}$el transition ($T_{N,Fe}$)), spin-reorientation transition of  $Fe^{3+}$ ions ($T_{SRT,Fe}$) and magnetic ordering of rare-earth ions \cite{RFOReview1, RFOReview2}. In almost all of $RFe0_3$ compounds, the weak ferromagnetic moment reorients from the orthorhombic $\textit{a}$ axis at low temperatures to the $\textit{c}$ axis at high temperatures. This temperature induced spin flop is known as spin-reorientation transition (SRT) i.e., a transition in which $Fe^{3+}$ spins reorient from one crystallographic axis to another. For example, $DyFeO_3$ exhibits paramagnetic to weak ferromagnetic transition corresponding to $Fe^{3+}$ at about 645 K,  SRT of $Fe^{3+}$ spins from canted AFM to collinear AFM state at about 62 K below which the weak ferromagnetism disappears and finally AFM ordering of $Dy^{3+}$ at low temperatures ($\approx$ 4 K) \cite{JAlCo}.  Similarly (GdFO) exhibits paramagnetic to weak ferromagnetic transition corresponding to $Fe^{3+}$ at about $T_{N,Fe}$ = 660 K and AFM ordering of $Gd^{3+}$ below 2.5 K \cite{Tokunaga, JAlCo}. It is to be noted that temperature induced SRT is not observed in GdFO, but rather it exhibits magnetic field induced SRT phenomena and the critical field required to induce SRT is found to be temperature dependent \cite{Durbin}.

Recently, Tokunaga et al., have reported that single crystal $GdFeO_3$ (GdFO) exhibits not only a weak ferromagnetism but also a ferroelectric ground state and demonstrated the mutual controllability of the two order parameters \cite{Tokunaga}. The appearance of FE ordering is explained in terms of the exchange interaction between the Gd and Fe spins. Similarly $DyFeO_3$ is another rare-earth ortho-ferrite which is recently reported to exhibit magnetic field induced ferroelectricity and  large ME coupling at about 4 K in single crystal form. The phenomena is explained in terms of exchange-striction between adjacent $Fe^{3+}$ and $Dy^{3+}$ layers with the respective layered antiferromagnetic (AFM) components \cite{TokungaPRL}. However, FE polarization is reported in the weak ferromagnetic region ($T_{SRT, Fe}$ $\leq$ T $\leq$ $T_{N,Fe}$) of polycrystalline $DyFeO_3$ without applying external magnetic field which suggests that the rare earth magnetic ordering is not essential for FE but weak ferromagnetism is sufficient \cite{Rajeswaran}. Similarly, spin-canting induced improper FE and spontaneous magnetization is reported in another rare-earth orthoferrite viz., $SmFeO_3$, whose origin is being debated \cite{SMFO1, SMFO2, SMFO3}. Therefore, the role of these magnetic transitions on the occurrence of FE ordering and hence multiferroic nature in these compounds is yet to be clearly established. 

M$\ddot{o}$ssbauer spectroscopy is a unique technique to study different magnetic transitions such as temperature and magnetic field induced SRT phenomena \cite{MossBook}. But most of the M$\ddot{o}$ssbauer literature deals with the study of single crystal $RFeO_3$ compounds, in which there exists a well defined angular relation between $Fe^{3+}$ spins and the incident gamma rays, either as a function of temperature or applied magnetic fields \cite{Durbin, Niklov}. It may be noted  because of the fact that there is a random distribution of $Fe^{3+}$ spins in polycrystals, there is a possibility of missing these kind of transitions when explored using high field M$\ddot{o}$ssbauer spectroscopy.  Probably because of this, the high field M$\ddot{o}$ssbauer spectroscopy has been used only for limited AFM polycrystals such as goethite ($\alpha$-FeOOH), haematite ($\alpha-Fe_2O_3$), barium ferrite etc., to study SRT phenomena \cite{LTHMMoss, Pankhurst1, Pankhurst2}. However, as most of recent literature on $RFeO_3$ compounds deal with the polycrystalline state of the samples \cite{JAlCo, Rajeswaran}, it would be required to explore the SRT phenomena in $RFeO_3$ polycrystals too using suitable techniques  and in the present work we show that low temperature high magnetic field M$\ddot{o}$ssbauer spectroscopy gives the signatures of SRT in polycrystalline $GdFeO_3$. 

Other aspect that is being recently reported is the role of spin-phonon coupling in stabilizing the FE ordering in similar compounds i.e., rare-earth chromites / manganites ($RCrO_3$ or $RMnO_3$). For example, Ferreira et al., reported spin phonon coupling as the origin of multiferroicity in $GdMnO_3$ \cite{Ferreira}. Mandal et al.,  reported the spin-phonon coupling in Mn doped $YFeO_3$ \cite{Mandal}. Bhadram et al., reported the Raman spectroscopy measurements of different orthochromites and concluded that the spin-phonon coupling is observed only for those cases in which the rare-earth ion is magnetic and concluded that spin-phonon coupling play important role in inducing FE ordering in $GdCrO_3$ type of compounds \cite{Bhadram}.  Therefore, study of phonons is expected to give the information regarding spin-phonon coupling, which has been argued to play important role in stabilizing the multiferroic ordering \cite{Ferreira, Mandal}. Temperature dependent Raman spectroscopy across the transition temperatures  is considered to be an effective tool to probe structural transitions, spin-phonon, spin-lattice coupling etc., in different materials \cite{Ferreira, Mandal, Bhadram}.  

In view of the above mentioned issues, in the present work we report the preparation of polycrystalline $GdFeO_3$ (GdFO) and the study of magnetic field induced SRT using low temperature high magnetic field $^{57}Fe$ M$\ddot{o}$ssbauer, spin-phonon coupling using temperature dependent Raman spectroscopy, bulk magnetization measurements.  

\begin{figure}[b]
  \centering
  \includegraphics[scale=1, width = 8 cm]{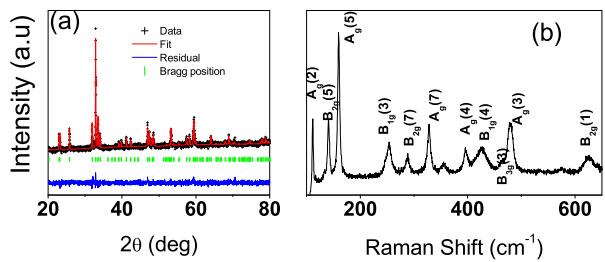}
  \caption{(a) X-ray diffraction pattern and (b) Raman spectrum with the indicated modes of polycrystalline $GdFeO_3$.}
\end{figure}

\maketitle \section{Experimental}
Polycrystalline $GdFeO_3$ (GdFO) is prepared with the conventional solid state route. X-ray diffraction (XRD) measurements are carried out using D8-Discover system of M/s Brucker equipped with Cu-K$\alpha$ radiation.  The dc-magnetization measurements were carried out using Quantum Design 14 Tesla  PPMS-VSM. The $^{57}Fe$ M$\ddot{o}$ssbauer measurements were carried out in transmission mode with a $^{57}Co$ (Rh) radioactive source in constant acceleration mode using a standard PC-based M$\ddot{o}$ssbauer spectrometer equipped with a WissEl velocity drive. Velocity calibration of the spectrometer was carried out with a natural iron absorber at room temperature. For the low temperature and high magnetic field measurements, the sample was placed inside a Janis superconducting magnet. The direction of the external field is parallel to gamma rays. For high temperature (300 - 683 K) M$\ddot{o}$ssbauer measurements, the sample, in argon gas flow, is placed inside a specially designed M$\ddot{o}$ssbauer furnace (MF-1100) with a temperature accuracy of about $1^o$C.  Raman spectra were recorded in back-scattering geometry using Jobin Yvon Horibra LABRAM spectrometer using He-Ne laser of 632.8nm as the excitation source. High temperature Raman measurements were carried out in THMS600 sample stage of Linkam Scientific instruments Ltd.

\begin{figure}[b]
  \centering
  \includegraphics[scale=1, width = 8 cm]{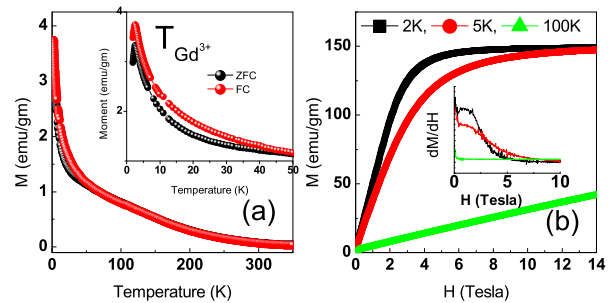}
  \caption{(a)Temperature dependent magnetization (M-T) data of polycrystalline $GdFeO_3$ measured under zero-field cooling (ZFC) and field cooling (FC) protocol in a field of 500 Oe. Inset show the enlarged view of M-T data depicting magnetic ordering of $Gd^{3+}$ ions. (b) M-H data of polycrystalline $GdFeO_3$ measured at different temperatures. Inset show the derivative of magnetization with respect to magnetic field.}
\end{figure}

\maketitle \section{Results and Discussions}

Figure-1(a) show the x-ray diffraction (XRD) pattern of the prepared GdFO sample along with Rietveld refinement. The XRD is fitted with Rietveld refinement, considering orthorhombic structure with $\textit{Pbnm}$ space group using FullProf software \cite{FullProf}. The XRD data confirms the single phase of the prepared sample and the refined lattice parameters viz., a=0.536 $\pm$ 0.001 nm, b=0.562 nm and c=0.768 nm, match well with literature \cite{JAlCo}. Figure-1(b) show the room temperature Raman spectrum of the prepared GdFO sample along with the identified Raman modes.  Experimentally observed Raman modes are assigned to the corresponding phonon mode with the help of lattice dynamical calculations reported earlier \cite{RamanLit1, RamanLit2}. According to the group theory, the irreducible representation for $RFeO_3$ compounds is given by  $\Gamma$= $7A_{g} + 7B_{1g} + 5B_{2g} + 5B_{3g} + 8A_{1u}  + 8B_{1u}  + 10B_{2u}  + 10B_{3u}$ in which 24 are Raman-active modes, 28 are infrared modes and 8 are inactive modes \cite{RamanLit1, RamanLit2, Bhadram}. The observed Raman modes of GdFO match closely with the recently reported Raman data of GdFO and also the other $RFeO_3$ compounds \cite{Du, RSC}.  After confirming the single phase of the prepared sample, detailed measurements viz., temperature and magnetic field dependent bulk magnetization, $^{57}Fe$ M$\ddot{o}$ssbauer and temperature dependent Raman spectroscopy are carried out on the prepared sample. 

Figure-2(a) show the temperature dependent magnetization (M-T) data of the sample measured in 500 Oe field under zero-field cooling (ZFC) and field cooling (FC) protocol.  One can clearly see the signatures of magnetic ordering of $Gd^{3+}$ ions at about 2.5 K as shown in the inset of figure-2(a). Figure-2(b) shows the M-H data measured at different temperatures. The data measured at 100 K is linear till the highest field i.e., 14 Tesla, which is typically the case for antiferromagnetic samples.  However, a change of curvature i.e., non-linear magnetization response is observed for the M-H data measured at 5 and 2 K.  One can understand these observations in terms of magnetic field induced spin re-orientation transition (SRT).  In literature, the spin-flop field is taken as the field at which a maximum appears in the derivative of M and H curves as shown in inset of Figure-2(b) \cite{SFField}. At 5 K the transition seems to occur over a wide range of fields as compared to 2 K and also the transition at 2 K is taking place relatively at lower fields as compared to 5 K data. It is to be noted that spin-flop transition field would depend upon strength of exchange interaction and anisotropy field.  The present results can be understood by realizing that at low temperatures (2 K) the magnetic ordering of $Gd^{3+}$ ions essentially amplifies the effect of applied magnetic field on the SRT phenomena. Further, to study the SRT phenomena and AFM-PM transition,  temperature and field dependent $^{57}Fe$ M$\ddot{o}$ssbauer measurements are carried out as discussed below.

M$\ddot{o}$ssbauer spectroscopy gives the information about the magnetic interactions like spin reorientation, spin canting, magnetic ordering, order-disorder transitions etc., unambiguously  \cite{MossBook}. In order to investigate the temperature dependence of the iron sublattice magnetization, a systematic  study  of $^{57}Fe$ M$\ddot{o}$ssbauer spectroscopy was carried out in the temperature (5-683 K) range and with external applied magnetic field at low temperatures. Figure-3 show the temperature dependent M$\ddot{o}$ssbauer spectra of the prepared GdFO sample.   M$\ddot{o}$ssbauer spectra show six line pattern even at room temperature, which is consistent with the fact that the $T_{N,Fe}$ of GdFO is reported to be around 660 K \cite{JAlCo}. The observed M$\ddot{o}$ssbauer data is similar to that of Eibschutz et al \cite{Eibschutz}., in which $^{57}Fe$ M$\ddot{o}$ssbauer spectra of several orthoferrites is reported.  The M$\ddot{o}$ssbauer data covering wide range of temperatures i.e., below and above $T_N$ is fitted by Eibschutz et al., considering only one site as orthoferrites are expected to have only one type crystallographic iron site \cite{Eibschutz}.  The same methodology is used in the present work.  The observed M$\ddot{o}$ssbauer spectra along with the fitting using NORMOS-SITE software at specified temperatures are shown in Figure-3. All the spectra up to 573 K are fitted with single six-line pattern. However between 623-673 K the data is well fitted with one sextet and one doublet indicating the co-existence of magnetic ordering and paramagnetic behavior, which might be associated with the critical fluctuations near the transition temperature \cite{Eibschutz}.  At 683 K the Mossbauer spectrum is fitted with only paramagnetic doublet indicating the N$\grave{e}$el transition of Fe-sublattice from antiferromagnetic ordering to paramagnetic ordering. Obtained values of hyperfine parameters like internal magnetic hyperfine field ($H_{int}$), isomer shift (IS), quadrupole splitting (QS) with the temperature match with the literature \cite{Eibschutz}. The obtained $H_{int}$ as a function of temperature is plotted in Figure-4.

\begin{figure}[htbp]
 \centering
 \includegraphics[scale=1, width = 8 cm]{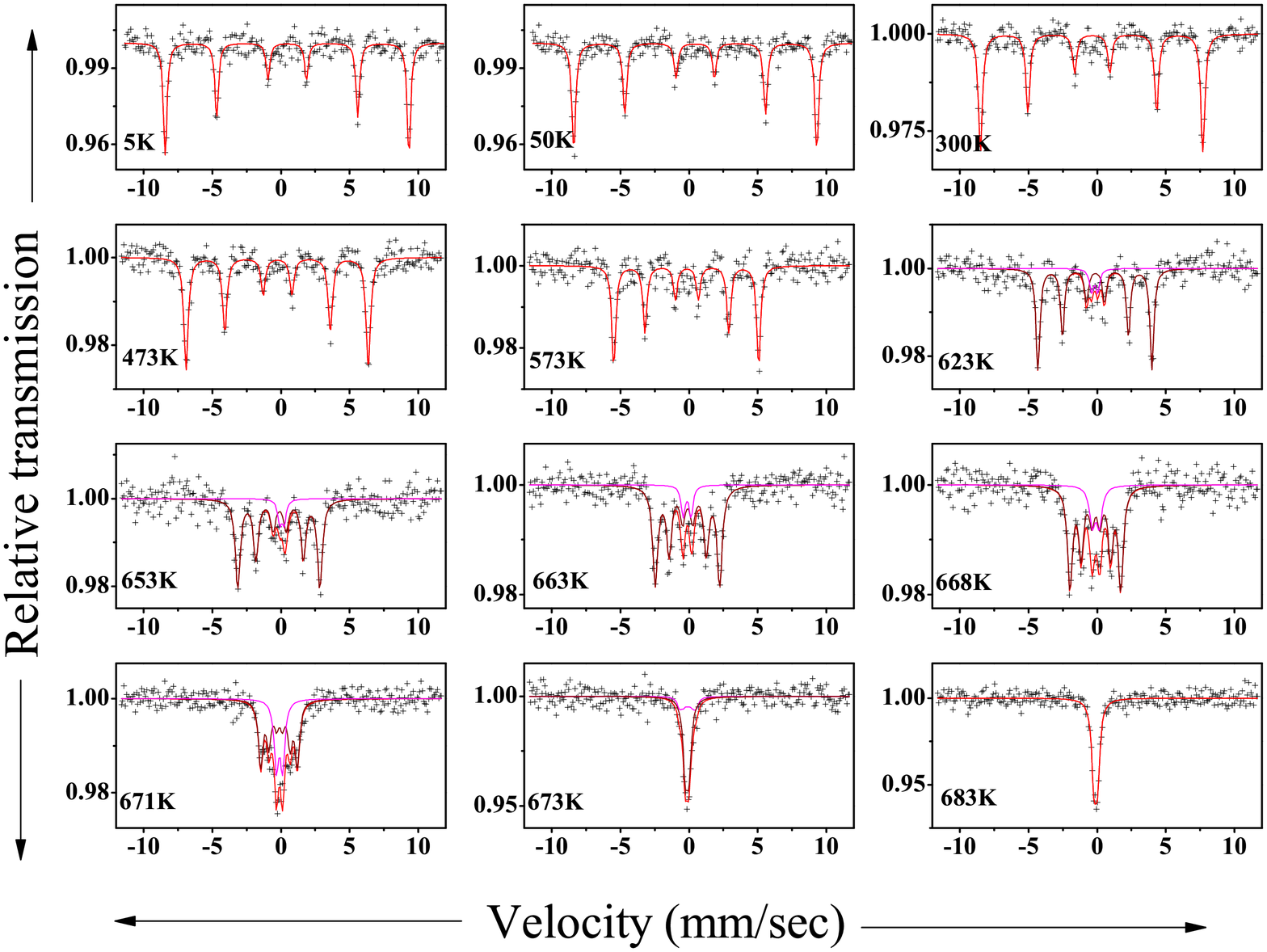}
\caption{Temperature dependent $^{57}Fe$ M$\ddot{o}$ssbauer spectra of $GdFeO_3$. Points are experimental data points and the solid line is the best fit to the data.}
\end{figure}

\begin{figure}[htbp]
  \centering
  \includegraphics[scale=1, width = 8 cm]{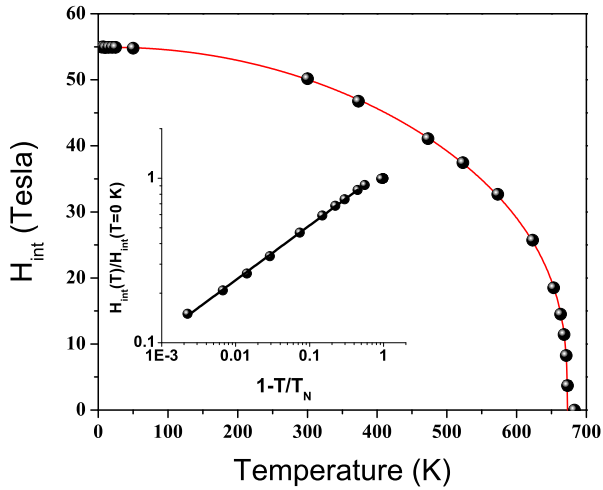}
  \caption{Variation of internal hyperfine filed ($H_{int}$) obtained from temperature dependent $^{57}Fe$ M$\ddot{o}$ssbauer spectra of $GdFeO_3$ as shown in Figure-3. Inset show the plot of $H_{int}$(T) / $H_{int}$($T_0$) versus 1-T/$T_N$ on a double logarithmic scale. Symbols represent the data and the solid line is the best fit to the data as described in the text. }
\end{figure}

For temperatures not sufficiently close to transition temperature, the temperature dependence of $H_{int}$ is given by $H_{int}$(T)/$H_{int}$($T_0$) = B $t^\beta$ [1+A $t^{\beta_1}$+O $t^{2\beta_1}$], where t=1-T/$T_N$, $\beta$ is the critical exponent for spontaneous magnetization, A is the correction-to-scaling amplitude and $\beta_1$ is the correction-to-scaling exponent \cite{Simopoulos}. The least square fit of the data as shown in Figure-4, resulted in the estimation of parameters viz., $T_N$ as 672.5$\pm$0.2 K and $\beta$ value of 0.315$\pm$0.002, $H_{int}$($T_0$) of 55.1 Tesla and $\beta_1$ is fixed at 0.55. The obtained value of $T_N$ matches reasonably with the reported value of GdFO \cite{JAlCo}.  However, the value of critical exponents are expected to be meaningful and accurate, if they are estimated from the region close to transition.  The temperature dependence of $H_{int}$ in the vicinity of transition temperature\cite{Eibschutz, BlaauwMoss} is often described by the formula $H_{int}$(T) $\propto$  $H_{int}$($T_0$) $t^\beta$. Plot of $H_{int}$(T) / $H_{int}$($T_0$) versus 1-T/$T_N$ on a double logarithmic scale yields a straight line as shown in the inset of Figure-4.  The straight line fit to the data in the region of 1-T/$T_N$ $\leq$ 0.55 gives the value of the critical exponent $\beta$ as 0.333$\pm$0.003.  The obtained $\beta$ indicates that the temperature dependence of internal hyperfine fields in GdFO follows the one-third law found in three-dimensional (3D) Heisenberg magnets. It is to be noted that the importance of estimation of critical exponents in multiferroic materials has recently been highlighted \cite{CriticalExpo1, CriticalExpo2, CriticalExpo3}. Since M$\ddot{o}$ssbauer measurements are basically zero external magnetic field measurements, the obtained value of $\beta$ is expected to be more realistic as compared to that obtained from bulk magnetization measurements \cite{CriticalExpo1, CriticalExpo2, CriticalExpo3}.

\begin{figure}[t]
\centering
\includegraphics[scale=1, width = 8 cm]{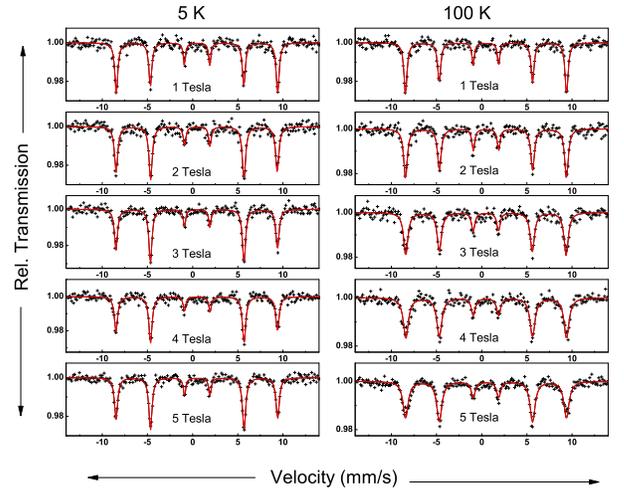}
\caption{$^{57}Fe$ M$\ddot{o}$ssbauer spectra measured at 5 and 100 K under the application of indicated external magnetic field. Points are experimental data points and the solid line is the best fit to the data.}
\end{figure}

\begin{figure}[b]
  \centering
  \includegraphics[scale=1, width = 8 cm]{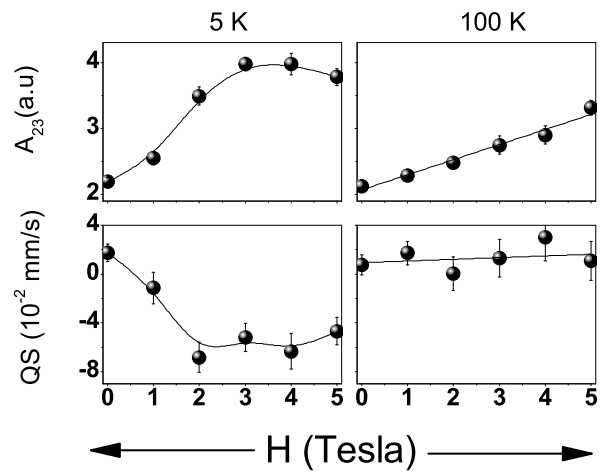}
  \caption{Variation of intensity of $\Delta$m=0 lines i.e., area ratio of second and third lines ($A_{23}$) and quadrupole splitting (QS) as a function of applied external field at 5 and 100 K as shown in Figure-5.  Solid line is guide to eye.}
\end{figure}

Figure-5 shows the $^{57}Fe$ M$\ddot{o}$ssbauer spectra measured at 5 and 100 K under the application of external magnetic field,  to study the field induced SRT phenomena.  M$\ddot{o}$ssbauer spectra contain information on the spin direction viz., angle between the $Fe^{3+}$ spins and the gamma-ray beam ($\theta$), which determines the relative intensity of the $\Delta$m=0 lines in the magnetic spectrum according to the relation 3: 4 $sin^2\theta$/(1 + $cos^2\theta$): 1 in thin absorber limit \cite{MossBook}. It is to be noted that AFM materials with the application of external magnetic field ($H_{ext}$) display a wide variety of phase transitions in addition to AFM-PM transition at the N$\grave{e}$el temperature.  If $H_{ext}$ is applied along the easy axis of magnetization of a uniaxial AFM below N$\grave{e}$el temperature, after a certain critical field depending on the anisotropy of the sample a first-order phase transition occurs and the spins rotate in a direction perpendicular to $H_{ext}$, which is known as spin-flop transition. With further increase of $H_{ext}$, the spins tip toward each other and exhibit a second-order phase transition to a phase (spin-parallel) in which the spins are parallel to the easy magnetization axis. The spin-flop transition is characterized by a sextet with an effective field of $\sqrt{(H_{ext}^2+H_{int}^2)}$ and $\Delta$m=0 lines with maximum intensity, whereas the spin-parallel phase is characterized by a sextet with an effective field of $H_{ext}$$\pm$$H_{int}$ and zero intensity of $\Delta$m=0 lines \cite{LTHMMoss}.  Many reports are published in single crystal $RFeO_3$ compounds, in which there exists a well defined angular relation between $Fe^{3+}$ spins and the incident gamma rays, either as a function of temperature or applied magnetic fields exhibiting the variation of intensity of $\Delta$m=0 lines taken as a finger print for the SRT phenomena \cite{Durbin}. 

In the present sample, one can also see that there is a variation in intensity of second and fifth lines of M$\ddot{o}$ssbauer sextet i.e., lines corresponding to $\Delta$m=0 transition as a function of applied external field (Figure-5).  This feature is often denoted by $A_{23}$ (area ratio of second and third lines) in a given sextet and is expected to be around 2.0 for polycrystalline sample corresponding to random spin orientation in zero applied magnetic field \cite{MossBook}. In view of the noticeable intensity variation of the $\Delta$m=0 lines as a function of applied magnetic field, the M$\ddot{o}$ssbauer data of the present study is fitted by keeping $A_{23}$ parameter as free variable and the obtained variation is shown in Figure-6. One can clearly see that $A_{23}$ variation resembles with that of M-H data as shown in figure-2(b). Quadrupole splitting (QS) is the other hyperfine parameter which gives information about SRT. In the magnetically ordered state, the presence of QS is usually treated as a small perturbation on the magnetically split nuclear energy levels given as $-3\textit{eQ}V_{ii}^H/4I(2I-1)$, where I=3/2, $\textit{e}$ is the electron charge, Q is nuclear quadrupole moment, $V_{ii}^H$ is the electric field gradient (EFG) component in the direction of quantization of nuclear moment i.e., the direction of $H_{int}$ \cite{MossBook, Eibschutz}. Therefore, for SRT in which $Fe^{3+}$ spins reorient from one crystallographic axis to another, the QS is expected to change sign.  This phenomena is observed in various single crystal $RFeO_3$ samples such as Ho doped $DyFeO_3$ \cite{Niklov}. In the present work also as shown in Figure-6, the variation of QS mimics the variation of $A_{23}$ indicating the field induced SRT in polycrystalline GdFO.

Therefore, the non-linear M-H data at 5 K is essentially due to the magnetic field induced SRT phenomena. The critical fields required to induce the SRT phenomena are quite different in polycrystalline GdFO as compared to single crystal GdFO as reported by Durbin et al \cite{Durbin}.  For the single crystal GdFO at 4.2 K the magnetic field required to induce SRT is about 1.2 Tesla, whereas magnetic fields of more than about 3 Tesla are required to induce SRT at 5 K as shown in the present work. Also the SRT with field at a given temperature is found to be broad as evident from dM/dH curves shown in Figure-2(b). One can understand these features in terms of the polycrystalline nature of the sample as $H_{ext}$ is differently oriented relative to crystal axes of different small single crystalline particles in polycrystals. In the case of single crystal AFM, the $H_{ext}$ is expected to be  perfectly aligned with the easy anisotropy axis and hence resulting in a sharp SRT at a critical field $H_{SF}$ approximately given by $\sqrt{2H_EH_A}$, where $H_E$ and $H_A$ are the exchange energy and anisotropy energy respectively. However, for polycrystalline samples one expects the angle ($\phi$) between the $H_{ext}$ and the easy axis is evenly distributed between 0 and $\pi$, there would be a distribution of $H_{SF}$ resulting in a broad transition \cite{Pankhurst1, Pankhurst2}. Another consequence of polycrystalline nature is the different value of critical field for SRT as compared to single crystals. Morup suggested that critical field for the spin-flop transition ($H_{SF}$) will be enhanced for antiferromagnetic microcrystals due to the presence of uncompensated moments \cite{Morup}. Recently, Kumar et al., studied the spin-flop transition in polycrystalline AFM samples considering an isotropic distribution of $\phi$ values and concluded that the polycrystalline sample has a range of spin-flop transitions at $H_{ext}$ cos$\phi$ = $H_{SF}$, which explains the higher values of $H_{ext}$ required for spin-flop as compared to single crystal AFM samples \cite{KumarKolkata}.

\begin{figure}[htbp]
\centering
\includegraphics[scale=1, width = 8 cm]{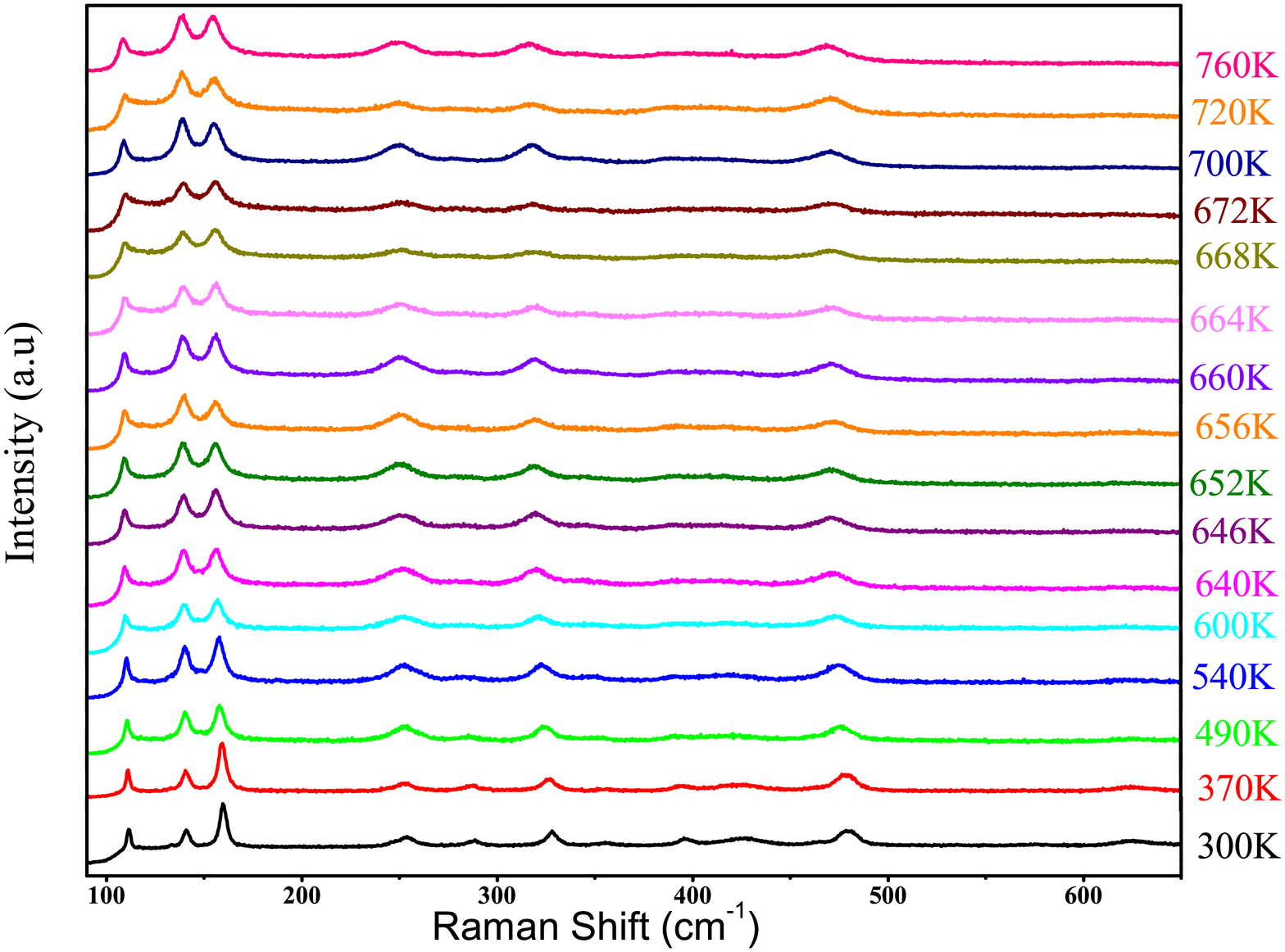}
\caption{Raman spectrum of polycrystalline $GdFeO_3$ measured at the indicated temperatures.}
\end{figure}

\begin{figure}[htbp]
  \centering
	  \includegraphics[scale=1, width = 8 cm]{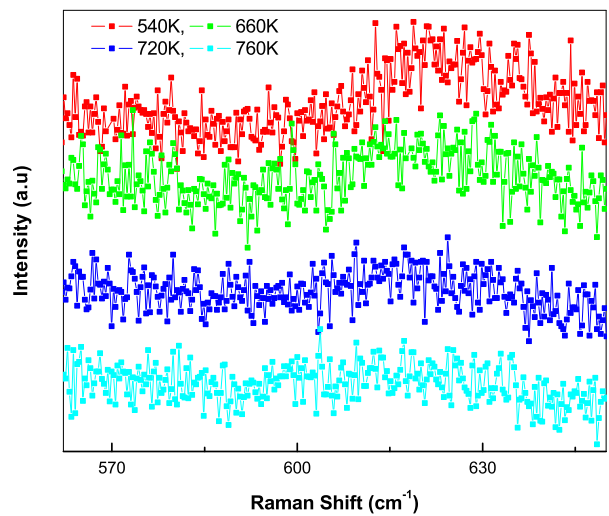}
	  \caption{Enlarged view of Raman data around $625 cm^{-1}$ at higher temperatures.}
\end{figure}

\begin{figure}[htbp]
  \centering
	  \includegraphics[scale=1, width = 8 cm]{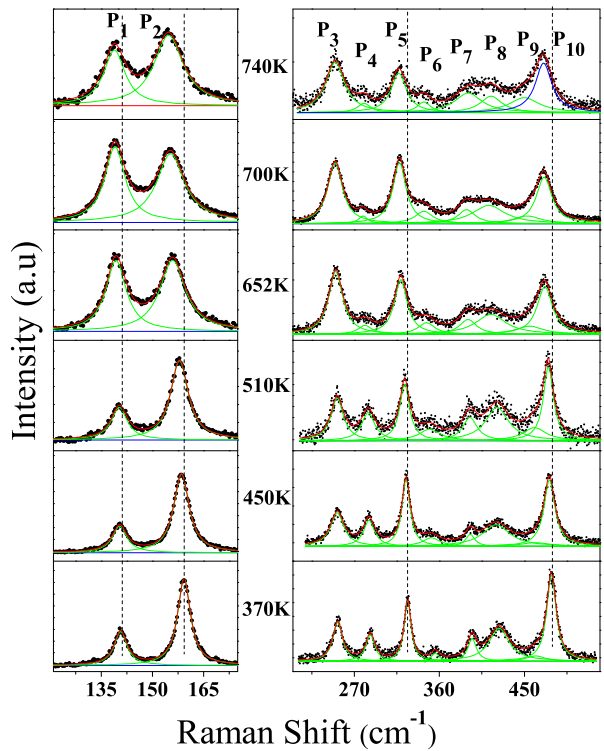}
	  \caption{Fitting of the Raman modes at representative temperatures. The vertical lines are guide to eye.  Obtained parameters viz., mode position, width, mode separation and area ratio as a function of temperature are plotted in Figure-10, 11 and 12.}
\end{figure}

Further, in order to study the behavior of phonons across the antiferromagnetic to paramagnetic transition (i.e., $T_{N,Fe}$) temperature dependent (300 - 760 K) Raman measurements are carried out and the data is shown in Figure-7. With increasing temperature, almost all the modes are found to shift to lower wave-number side and becoming broad consistent with thermal expansion.  It may appear that some of the modes with weak intensity are disappearing at higher temperatures, but essentially with increasing temperature the width is increasing and are merging with the background.  For example as shown in Figure-8, the mode at 625 $cm^{-1}$, which corresponds to in-phase stretching vibrational mode of $FeO_6$, is found to exist even at highest temperature studied.  Therefore, one can mention from Figure-7 that the overall Raman spectral signature is maintained across $T_{N,Fe}$, which is consistent with the fact that the AFM to PM transition is not accompanied by any structural phase transition.  $RFeO_3$ compounds undergo a structural transformation of orthorhombic - rhombohedral -cubic at much higher temperatures as compared to $T_{N,Fe}$ \cite{Neutron, HTXRD1, HTXRD2, HTXRD3}. For example $LaFeO_3$ with $T_{N,Fe}$ of about 735 K undergoes the orthorhombic - rhombohedral transition at 1228 K and the rhombohedral -cubic transition is expected about 2140 K \cite{HTXRD1, HTXRD2, HTXRD3}. 

However, a closer inspection of the temperature dependent Raman spectra reveals interesting spectral changes as a function of temperature and especially in the vicinity of $T_{N,Fe}$. The evolution of the Raman spectra with temperature in the region of interest is de-convoluted into different peaks as shown in Figure-9. From this fitting, the obtained position and width of the selected Raman modes is shown in Figure-10.  It is observed that the magnitude of shift is different for different modes.  For example, the mode at about 480 $cm^{-1}$ show large shifts (of the order of about 10 $cm^{-1}$) from above $T_{N,Fe}$ to room temperature as compared to other modes. The mode at about 140 $cm^{-1}$ show a shift of about only 3 $cm^{-1}$.  One can understand this by realizing that in $RFeO_3$ compounds, it is reported that rare-earth atom vibrations play a dominant role in determining the phonon frequencies below 200 $cm^{-1}$, the transition metal ions play dominant  role above 350-400 $cm^{-1}$ and vibrations involving rare-earth and oxygen contribute in the intermediate region \cite{Bhadram, SSCRaman}. With temperature change across $T_{N,Fe}$, there can be either some volume change or contribution of spin energy to the magnetic ion ($Fe^{3+}$) displacement  resulting in large shifts \cite{SSCRaman}. 

\begin{figure}[t]
  \centering
  \includegraphics[scale=1, width = 8 cm]{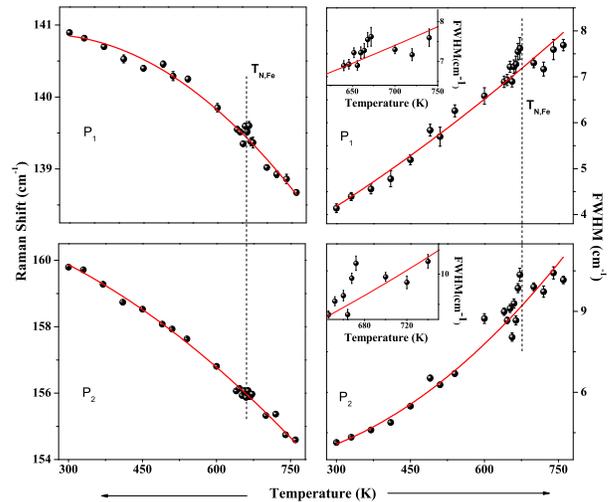}
  \caption{Temperature variation of position and width of different Raman modes as indicated in Figure-9. Symbols are the data points and the solid line is the best fit to the data considering the anharmonic model given by Balkanski et al \cite{Balkanski}. Vertical dotted line is to show the N$\grave{e}$el temperature ($T_{N,Fe}$). Inset is the enlarged view near $T_{N,Fe}$. }
\end{figure}

\begin{figure}[htbp]
  \centering
  \includegraphics[scale=1, width = 8 cm]{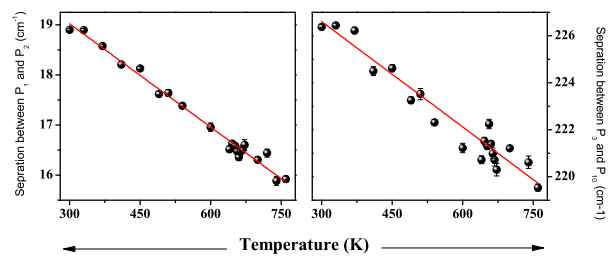}
  \caption{Separation of Raman modes as a function of temperature for (a) P1 and P2 (b) P3 and P10 of Figure-9. Symbols are data points and the solid line is the linear fit.}
\end{figure}

\begin{figure}[htbp]
  \centering
  \includegraphics[scale=1, width = 8 cm]{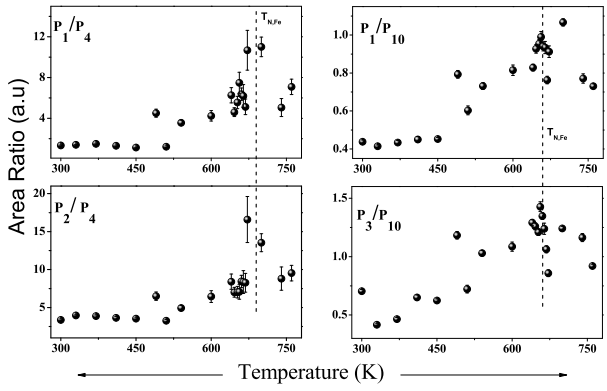}
  \caption{Area ratio of different Raman modes as a function of temperature.  P denotes the peak number as shown in Figure-9. Vertical dotted line is to show the N$\grave{e}$el temperature ($T_{N,Fe}$).}
\end{figure}

The behavior of the phonon position and phonon line-width with temperature follows the anharmonic model given by Balkanski \cite{Balkanski}.  The equation  $\Gamma$ (T) = A [1+ $\frac{2}{e^x-1}$] +B [1+ $\frac{3}{e^y-1}$ + $\frac{3}{(e^y-1)^2}$], where x=$\hbar\omega_0$/2$k_B$T, y=$\hbar\omega_0$/3$k_B$T, $\omega_0$, A and B are fitting parameters, which takes into account of the four phonon process is considered for fitting the obtained temperature variation of mode position and width as shown in Figure-10. One can see that the data is fitted well by this model. However, for some of the modes anomaly in mode width and also subtle changes in mode position are observed across $T_{N,Fe}$.  As phonon life time (or lattice relaxation time) and line-width obey the  uncertainty relation between time and energy so this anomaly near $T_{N,Fe}$ in the width of 140 $cm^{-1}$ can be regarded as spin-lattice coupling.  Similarly anomaly near $T_{N,Fe}$ in the phonon position  of 140 $cm^{-1}$ can be regarded as spin-phonon coupling.  As mentioned above, modes at lower wave-number (140 $cm^{-1}$) corresponds to the movement of $Gd^{3+}$ ions.  Therefore, it is interesting to note that the correlation between the $Gd^{3+}$ ions and the $Fe^{3+}$ spin ordering exists even up to $T_{N,Fe}$.  This may throw some light on the controversy whether it is the rare-earth magnetic ordering (which takes place at temperatures below 5 K) or the weak ferromagnetic state of $Fe^{3+}$ (which takes place below $T_{N,Fe}$) ions play important role in stabilizing the ferroelectric ordering in these type of compounds.  

Another interesting observation from the temperature dependent Raman data is that, one can clearly see the systematic change in the intensity of modes and also the separation of modes with temperature (Figure-9). Figure-11 shows the the separation of modes at  about 140, 160 $cm^{-1}$ (denoted as P1 and P2 in  Figure-9) and at about 250, 459 $cm^{-1}$ (denoted as P3 and P10 in  Figure-9) as a function of temperature. With increasing temperature, for both the cases the separation is found to decrease indicating the shrinking of the spectra. This is essentially due to the fact that some of the modes are shifting at a much faster rate with temperature as compared to the remaining indicating anisotropic nature of the force constants between the atoms.  Also the area ratio of different modes as shown in Figure-12 exhibits an anomalous behavior close to $T_{N,Fe}$. It is to be noted that the area of a Raman mode apart from its position and width is considered to be an effective parameter to study the vibrational dynamics in a given material.  Therefore, the observed anomalous variation of area ratio of different Raman modes in the present work unambiguously indicate the presence of spin-lattice coupling in polycrystalline $GdFeO_3$.  In addition to this, as the observed Raman modes correspond to vibrational motion of the rare-earth and iron atoms in different directions, the observed anomalous variation close to $T_{N,Fe}$ highlight the role of magnetic ordering in inducing local structural re-arrangement which might be responsible for the observation of ferroelectric and hence multiferroic nature in these type of compounds.  Recently Selbach et al., reported that there is an anisotropic thermal expansion of lattice parameters in $LaFeO_3$ even across $T_{N,Fe}$ and proposed that magneto-striction could be the reason \cite{HTXRD1}.  Detailed temperature dependent x-ray diffraction and fine structure measurements might help in corroborating the present Raman observations of polycrystalline GdFO.      

In conclusion, the M$\ddot{o}$ssbauer spectra measured in the presence of external magnetic field show the signatures of field induced spin reorientation transition in polycrystalline $GdFeO_3$, which are corroborated by magnetization measurements. From the temperature dependent variation of internal hyperfine field, N$\grave{e}$el transition temperature ($T_N$) of 672.5$\pm$0.2 K and critical exponent ($\beta$) of 0.333$\pm$0.003 is obtained. Temperature dependent (300 - 760 K) Raman spectroscopy measurements show the signatures of spin-phonon coupling and local structural re-arrangement across $T_{N,Fe}$.       
 
\maketitle\section {Acknowledgments}
Thanks are due to Dr.M.Gupta and Mr.L.Behra for XRD; Dr.Alok Banerjee and Mr.Kranti Kumar for magnetization measurements .  

\bibliography{manuscript}

\end{document}